\documentclass[aps,prl,floatfix,twocolumn,footinbib,amsmath,amssymb]{revtex4}
\usepackage{amsmath,amsthm,amssymb,color,psfrag}
\usepackage{url}
\usepackage{latexsym}
\usepackage{graphicx}
\usepackage{slashed}

\definecolor{darkred}{rgb}{0.6,0.0,0.0}
\definecolor{darkblue}{rgb}{0.0,0.0,0.5}
\definecolor{darkgreen}{rgb}{0.0,0.5,0.0}
\definecolor{brown}{rgb}{0.0,0.0,0.0}

\begin{document}

\topmargin 0.0in

\title{Telescoping Jets: Multiple Event Interpretations with Multiple $R$'s}
\author{Yang-Ting Chien}
\affiliation{
Center for the Fundamental Laws of Nature,
Harvard University,
Cambridge, MA 02138, USA}

\begin{abstract}
Jets at high energy colliders are complicated objects to identify. 
Even if jets are widely separated, 
there is no reason for jets to have the same size.
A single reconstruction, or {\sl interpretation}, of each event can only extract a limited amount of information. Motivated by the recently proposed Qjet algorithms, which give multiple interpretations for each event using nondeterministic jet clustering, we propose a simple, fast and powerful method to give multiple event interpretations by varying the parameter $R$ in the jet definition. With multiple interpretations we can redefine the weight of each event in a counting experiment to be the fraction of interpretations passing the experimental cuts, instead of 0 or 1 in a conventional analysis. We show that the statistical power of an analysis can be dramatically increased. In particular, we can have a $46\%$ improvement in the statistical significance for the Higgs search with an associated {\sl Z} boson ($ZH\rightarrow \nu {\bar \nu}b {\bar b}$) at the 8 TeV LHC. 
\end{abstract}
\maketitle

Jets are manifestations of the underlying colored partons in hard scattering processes. In order to
reconstruct 
hard processes and uncover physics at high energy, 
jets are key objects to identify in high energy collider experiments. The conventional way to identify jets is to use clustering algorithms 
\cite{Cacciari:2008gp,Dokshitzer:1997in,Wobisch:1998wt,Catani:1993hr,Ellis:1993tq},
\begin{figure}
\includegraphics[width=0.28\textwidth]{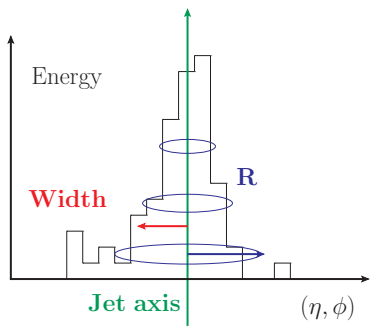}
\caption{\label{fig:R}A cartoon calorimeter plot distinguishing the width of the localized energy distribution of a jet (red) from the parameter $R$ (blue) in the anti-$k_T$ algorithm. $R$ is an artificial distance scale introduced to define the calorimeter region we want 
to look at. The jet axis points in the direction of the dominant energy flow, and the precise direction is not essential here.}
\end{figure}
where a parameter $R$ sets an {\sl artificial} jet 
size. The constituents of each reconstructed jet are 
those particles within an angular scale $R$ away from the jet direction. This is particularly true for the anti-$k_T$ algorithm because it gives almost perfect cone jets
in the calorimeter pseudorapidity-azimuthal angle ($\eta$-$\phi$) plane.
On the other hand, a jet 
is a distinct structure in its own right
with many 
collinear particles.
The width of the localized energy distribution of the jet in the $\eta$-$\phi$ plane is 
an independent quantity and should be distinguished from the parameter $R$ 
(FIG. \ref{fig:R}).

\begin{figure}[b]
\includegraphics[width=0.41\textwidth]{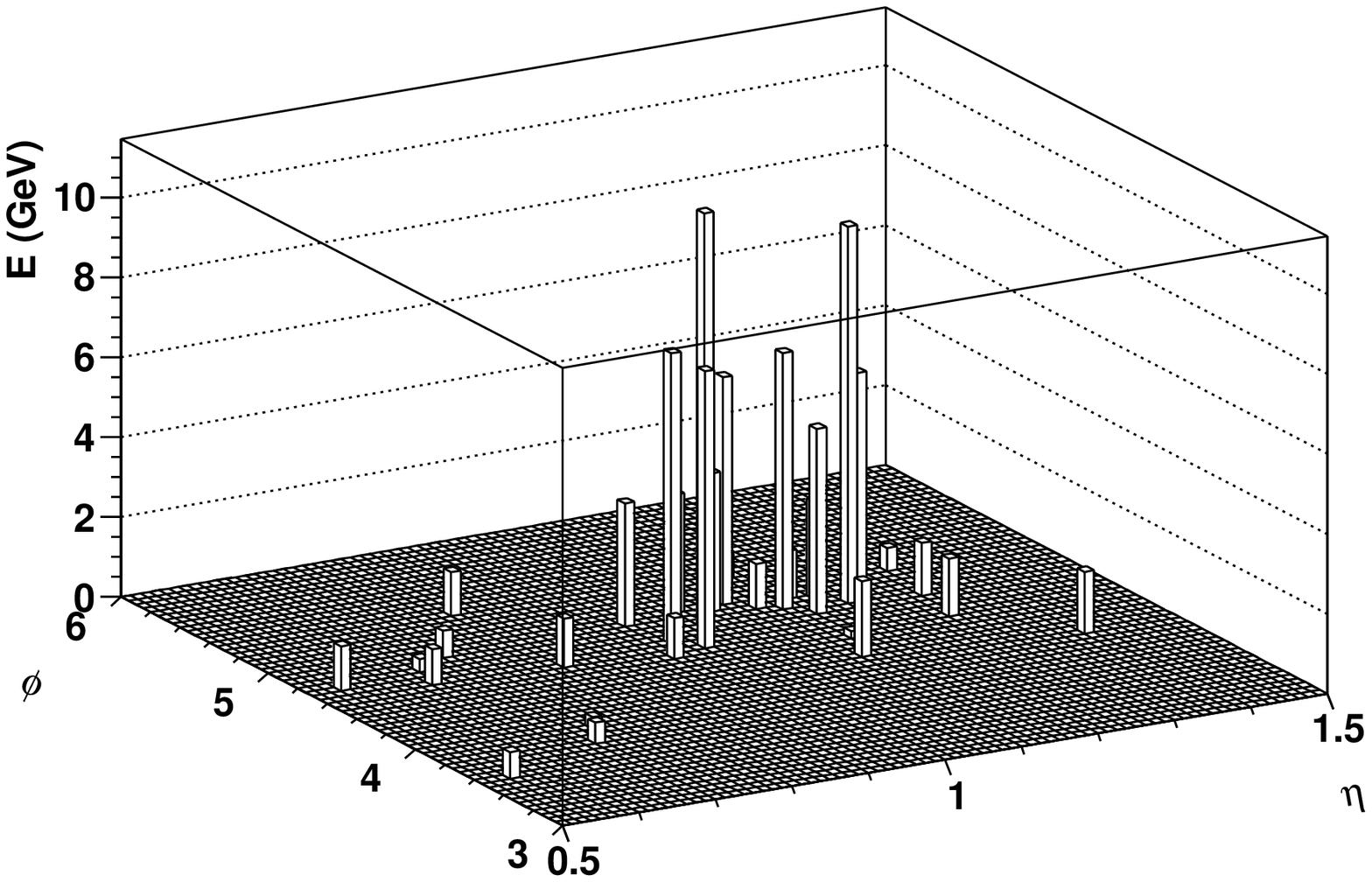}
\includegraphics[width=0.41\textwidth]{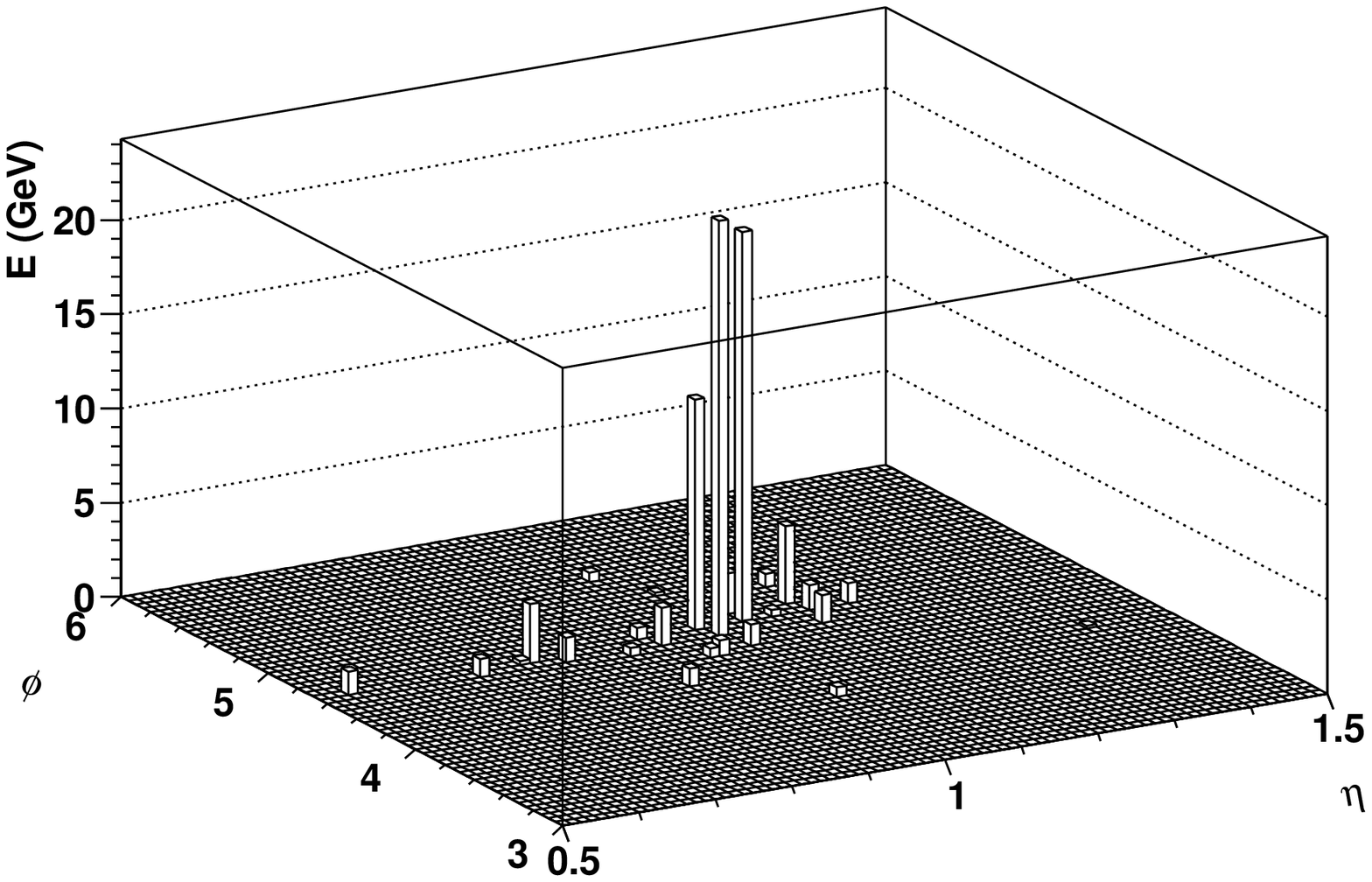}
\caption{\label{fig:Rs} Two {\sl b} jets with the same partonic kinematics but different widths, wider (top) and narrower (bottom).}
\end{figure}

Because the formation of jets 
is quantum mechanical and probabilistic,
the widths of jets 
are always different (FIG. \ref{fig:Rs}). To reconstruct partonic kinematics
we should pick a large enough $R$ so that most of the radiation emitted by the partons is enclosed. However, 
with a large $R$ more radiation contamination will be included. 
We can
manage to use jet grooming techniques \cite{Butterworth:2008iy,Ellis:2009me,Ellis:2009su,Krohn:2009th} 
to get rid of contamination. 
Algorithms with a large $R$ may also fail to resolve jets 
in some events.
Multiple partons 
may be 
in a fat jet which potentially has substructure. Without looking into jet substructure we may incorrectly 
include irrelevant jets in event reconstruction. 
In the end an $R$ is chosen for all events to optimize an analysis (see \cite{Krohn:2009zg} for jets with variable $R$).
A fixed $R$ defines a single set of constituents for each jet and a single {\sl interpretation} for each event. There is no choice of $R$ in conventional clustering algorithms which can 
resolve jets and get 
most of the relevant radiation for all events.

Multiple event interpretations can provide extra information and help increase the statistical power of an analysis. The recently proposed Qjet algorithms \cite{Ellis:2012sn} give multiple event interpretations using nondeterministic jet clustering. Unlike conventional clustering algorithms, Qjets merge pairs of particles probabilistically according to an exponential weight, resulting in different clustering histories.
An event may have a wide range of interpretations, and the probabilistic nature of Qjets allows the correct event structure
to emerge. 
It was shown that jet sampling with Qjets \cite{Kahawala:2013sba} can help improve considerably 
in the statistical significance $S/\delta B$ --the expected size of the signal divided by the background uncertainty-- in many classes of analyses, and it is interesting to understand the essence of Qjets.

\begin{figure}[t]
\includegraphics[width=0.35\textwidth]{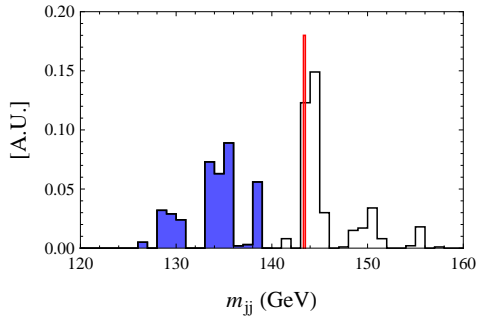}
\caption{\label{fig:distribution} The invariant mass distribution of the two {\sl b} jets for a $ZH$ event with multiple interpretations using the
telescoping jet algorithms (black). Using the anti-$k_T$ algorithm with $R$=0.7, $m_{jj}$=143.4 GeV (red) which is 
outside the mass window of 110 GeV $<m_{jj}<$ 140 GeV in a conventional analysis. Using multiple interpretations reveals the ambiguity of this event and $37\%$ of the interpretations pass the cuts (blue).}
\end{figure}


In this paper we propose a 
simple way 
to define an event interpretation: each choice of $R$ in jet algorithms gives a distinct event interpretation. The idea of probing jets with multiple $R$'s is referred to as {\bf telescoping jets}.
As a first step we can apply conventional clustering algorithms on each event multiple times with different $R$'s. Note that, with a too-small $R$ we may resolve an event in too much detail that miss its overall jet structure: in the $R\rightarrow0$ limit particles are all jets. 
On the other hand, with a too-large $R$ we may fail to resolve close jets. 
To deal with these issues, we improve 
the algorithm by first using the anti-$k_T$ algorithm with a suitable 
$R$ to reliably reveal the jet structure of an event and determine the jet axes from the reconstructed jet "cores". These axes point in the directions of the dominant energy flow in an event, and the precise directions are not essential. 
We can also use the axes determined through a jet shape minimization procedure 
and bypass using clustering algorithms. Then we define jet constituents by the particles within a distance $R$ away from the predetermined jet axes in the $\eta$-$\phi$ plane. So different interpretations correspond to different jet constituents without the tree structure.

However, another way of thinking about the above 
telescoping cone algorithm is that, we essentially move {\sl down} the clustering sequence in the anti-$k_T$ algorithm to build up jets after identifying the branch structure. This is complimentary to moving {\sl up} the reclustered 
tree and looking for mass drops to identify the branches \cite{Butterworth:2008iy,Kaplan:2008ie}. Using different $R$'s allows us to probe the energy distribution within each jet and give multiple event interpretations, and every observable of each event turns from a single number to a distribution (FIG. \ref{fig:distribution}).

\begin{figure}[t]
\includegraphics[width=7cm]{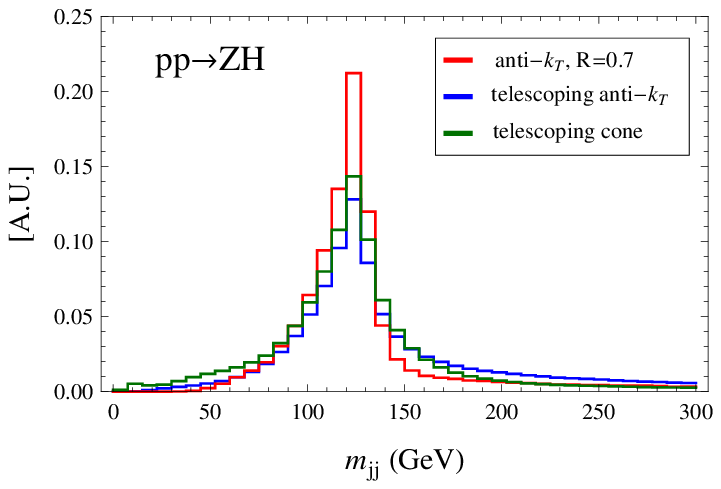}
\includegraphics[width=7cm]{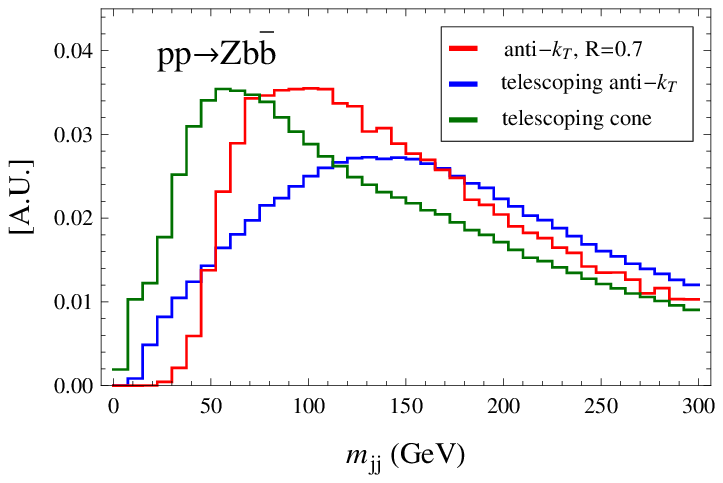}
\caption{\label{fig:mass}The signal (top) and background (bottom) $m_{jj}$ distributions reconstructed using the anti-$k_T$ algorithm with $R$=0.7 (red), as well as the telescoping anti-$k_T$ (blue) and cone (green) algorithms. 
Using multiple event interpretations gives a wider signal Higgs mass peak, but it reduces the statistical fluctuations of the $m_{jj}$ distributions. }
\end{figure}

In the following we present the detailed procedure of the algorithm
and apply it in a 
search for associated production of a Higgs and a {\sl Z} where the Higgs decays to two {\sl b} jets and the {\sl Z} decays to $\nu\bar\nu$ ($ZH\rightarrow \nu\bar\nu b \bar b$). The background is $Z+b\bar b$ 
from $g\rightarrow b \bar b$. 
We require the events to pass a ${\slashed E}_T > 120$ GeV cut for the experimentally available triggers. 
The $b \bar b$ system is slightly boosted so that the two {\sl b} jets are closer to each other 
and more difficult to resolve.
We define the signal window (specified later) by imposing cuts on the invariant mass of the two {\sl b} jets $m_{jj}$ (FIG. \ref{fig:mass}) and the transverse momentum of each {\sl b}-jet in our analysis. With multiple interpretations, each event is counted by the fraction of interpretations passing the cuts, instead of 0 or 1 in a conventional analysis. As we will see, this increases the statistical stability of observables so that background fluctuations shrink considerably, which is the key for $S/\delta B$ improvement.


\begin{figure}[t]
\includegraphics[width=0.45\textwidth]{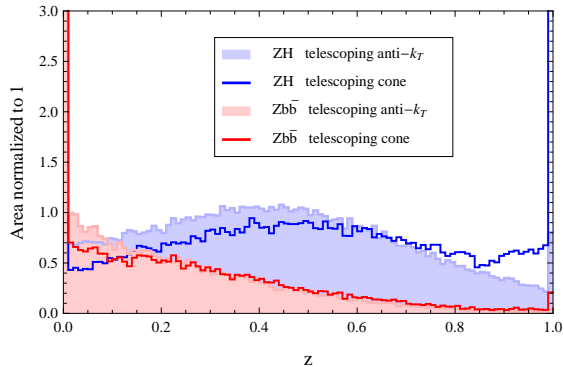}
\caption{\label{fig:efficiency}The signal and background $z$ distributions $\rho_S(z)$ and $\rho_B(z)$ using the telescoping anti-$k_T$ 
and cone algorithms. $z$ is the fraction of interpretations of an event passing the experimental cuts. A large fraction of both signal and background events can be interpreted differently.}
\end{figure}


In the context of Higgs search in the $ZH\rightarrow \nu\bar\nu b \bar b$ channel, we first use the telescoping anti-$k_T$ algorithm to reconstruct the two hardest jets with {\sl N} different $R$s, giving $N$ interpretations for each event.
The scaled-up computation time is 
tiny compared to using nondeterministic clustering algorithms \cite{Kahawala:2013sba}.
Here we take $N$=100. 
The value of $R$ ranges from 0.2 to 1.5 with an increment $\frac{1.3}{N}$. The range of $R$ is chosen because 
with the ${\slashed E}_T > 120$ GeV cut the angular separation between the two {\sl b} jets will be roughly $\lesssim2$. Here each interpretation is weighted uniformly for simplicity.

The telescoping cone algorithm which captures the jet structure more correctly goes as follows:
\begin{itemize}
\item Use the anti-$k_T$ algorithm with $R$=0.4 to reconstruct the cores of the two hardest jets and determine the jet axes $n_1$ and $n_2$.
\item Define the {\sl i}-th jet to be the particles within 
a distance $R$ away from $n_i$ in the $\eta$-$\phi$ plane:
\begin{equation}
    ~~~~~~~~{\rm jet}_R^i=\{~p~|~(\eta_p-\eta_{n_i})^2+(\phi_p-\phi_{n_i})^2<R^2\}.
\end{equation}
\item In the case of overlapping jets, assign particles to the jet with the closer jet axis. This step is to avoid ambiguity and is not crucial when reconstructing the invariant mass of the two hardest jets $m_{jj}$.
\end{itemize}
Here we use the same $R$ for both {\sl b} jets in an event.
However, for generic beyond the standard model physics searches with both quark and gluon jets in the final state, one can exploit the full idea of using different $R$'s for different jets.
We will leave these for future studies.

Our signal and background events were generated at the parton level using Madgraph 5 \cite{Alwall:2011uj} and then showered with Pythia 6.4 \cite{Sjostrand:2006za} 
for the 8 TeV LHC.
We impose the ${\slashed E}_T > 120$ GeV cut at the Madgraph level and the following cuts in the analysis to define the signal window:
\begin{itemize}
\item 110 GeV $<m_{jj}<$ 140 GeV
\item Both $p_T$s of the two hardest jets $>$ 25 GeV.
\end{itemize}
We use the anti-$k_T$ algorithm implemented in Fastjet v3.0.0 \cite{Cacciari:2011ma,fastjet}, and we perform the 
analysis with $R$ at the optimized value of $R$=0.7. 
We then study how the statistical significance of the Higgs search changes using multiple event interpretations. With $N$ event interpretations $m_{jj}$ turns from a single number to a distribution for each event. We define $z$ to be the fraction of event interpretations passing the above cuts. FIG. \ref{fig:efficiency} shows the $z$ distributions $\rho_S(z)$ and $\rho_B(z)$ for signal and background.
This is in contrast to the conventional analysis in which an event either passes the cuts or does not. With multiple event interpretations we can gain more information about the degree of certainty of an event being signal-like. 
Weighting each event by $z$ in the counting experiment helps improve the significance of the analysis.

\begin{figure}[t]
\includegraphics[width=0.4\textwidth]{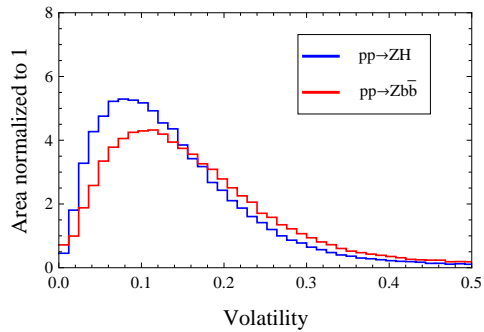}
\caption{\label{fig:vola}The signal (blue) and background (red) volatility distributions using 
the telescoping cone algorithm. }
\end{figure}

Let $\epsilon$ and $\sigma^2$ be the mean and variance of the $z$ distribution, and $N_S$ and $N_B$ be the expected numbers of signal and background events produced at the 8 TeV LHC. Then the significance is equal to
\begin{equation}
    \frac{S}{\delta B}=\frac{N_S ~\epsilon_S}{\sqrt{N_B(\epsilon_B^2+\sigma^2_B)}}.
\end{equation}
A more detailed discussion about statistics can be found in \cite{Kahawala:2013sba,Ellis:2013}. 
The {\sl volatility} (FIG. \ref{fig:vola}) of each event is defined by ${\cal V}$=$\Gamma/\langle m \rangle$,
where $\Gamma$ and $\langle m \rangle$ are the standard deviation and mean of the $m_{jj}$ distribution of each event with multiple interpretations. Volatility is useful 
in distinguishing boosted $W$ jets from their QCD background \cite{Ellis:2012sn}, and we will leave exploiting volatility in Higgs searches for future studies.


The performances of the algorithms are summarized in TABLE \ref{tab}.
\begin{table}[t]
\begin{tabular}{ccccc}
\hline
$R$ range & ~~~$N$~~~ &  algorithm & ~~~weight~~~ & $S/\delta B \uparrow$ \\
\hline
0.4 and 1.0& 2 & cone & $z$ & $14\%$\\
0.4 to 1.0& 7  & cone & $z$ & $20\%$\\
0.4 to 1.5& 12  & cone & $z$ & $26\%$\\
\hline
0.2 to 1.5& 100 & anti-$k_T$ & $z$ & $20\%$\\
0.2 to 1.5& 100 & cone & $z$ & $\bf 28\%$\\
\hline
0.4 to 1.5& 12  & cone & $\rho_S/\rho_B$ & $38\%$\\
0.2 to 1.5& 100 & cone & $\rho_S/\rho_B$ & $\bf 46\%$\\
\hline
\end{tabular}
\caption{\label{tab}$S/\delta B$ improvements using telescoping jets with different ranges of $R$, numbers of interpretations $N$, jet algorithms and weights in the counting experiment. }
\end{table}
The key for the $S/\delta B$ improvement is the shrink of background fluctuations, which comes from the rapid decrease of $\sigma_B$. For 
experimental studies with 
jet energy calibration depending on the parameter $R$, we try different ranges of $R$'s and fewer interpretations using the telescoping cone algorithm. 
Note that we can get half the improvement by using just two $R$'s, 
and using 12 $R$'s between 0.4 and 1.5 performs almost as good as using 100 $R$'s between 0.2 and 1.5.

With $\rho_S(z)$ and $\rho_B(z)$ we can get an even larger improvement 
with the optimized weight $\frac{\rho_S(z)}{\rho_B(z)}$ \cite{Kahawala:2013sba} 
in the counting experiment. 
Then the significance is equal to
\begin{equation}
    \frac{S}{\delta B}=\frac{N_S}{\sqrt{N_B}}\sqrt{\int_0^1\frac{\rho_S^2(z)}{\rho_B(z)}dz}~~,
\end{equation}
and we get a $46\%$ improvement compared to the conventional analysis. For $R$=0.4 to 1.5 with increment 0.1 we can get a $38\%$ improvement with just 12 $R$'s.

To conclude, 
the width of the localized energy distribution of a jet may not match well with the parameter $R$ in jet algorithms. The situation is even more complicated for events with close jets because resolving jets becomes an issue when the parameter $R$ and the distance between jets confront with each other.
We explore a simple and promising way 
of giving multiple interpretations for each event by changing the parameter $R$ in jet algorithms. 
The approach increases the statistical stabilities of observables which leads to remarkable improvement in the significance of a refined counting experiment.
Telescoping jets open up the possibility of refining and improving jet physics analysis in high energy experiments.

Also, we only look at the transverse momenta and invariant mass of the two {\sl b} jets, which are observables at high energy scales. It would be interesting to see how much more we can improve the significance of Higgs searches in hadronic channels by combining the analysis with other jet substructure \cite{Ellis:2010rwa,Stewart:2010tn,Thaler:2010tr} and color flow \cite{Gallicchio:2010sw,Hook:2011cq} observables, which probe softer sectors of QCD and color connections in an event. The approach of using multiple event interpretations could potentially be combined with likelihood ratio test and multivariate analysis, and in the presence of pile up our method will have to combine with jet grooming techniques. Applications of telescoping jets beyond physics searches, for example observable measurements, are also worth investigating. Probing jets with multiple $R$s may also allow us to 
construct jet observables more reliably. 

The author would like to thank David Farhi, Marat Freytsis, Dilani Kahawala, David Krohn, Matthew Schwartz and Jessie Shelton for 
helpful discussions and comments on the manuscript. 
The work is supported by DOE grant DE-SC003916. All the computations were performed on the Odyssey cluster at Harvard University.

\end{document}